\documentclass[12pt,letterpaper]{article}
\usepackage{amsmath}
\usepackage{amssymb}
\usepackage[dvips]{graphicx}
\usepackage{setspace}
\usepackage{enumerate}
\usepackage{hyperref}

\begin{document}

\onehalfspacing

\title{Hopping charge transport in disordered organic materials: where is the disorder?}
\author{S.V. Novikov and A.V. Vannikov\\
A.N. Frumkin Institute of Physical Chemistry and Electrochemistry,
\\ Leninsky prosp. 31, Moscow 119991, Russia}

\date{}

\maketitle

\begin{center}
\textbf{\large Abstract}
\end{center}

\noindent Effect of energetic disorder on charge carrier transport in
organic materials has been reexamined. A reliable method for
mobility calculation and subsequent evaluation of relevant
disorder parameters has been discussed. This method is well suited
for a direct calculation of the magnitude of dipolar disorder
$\sigma_\textrm{dip}$ in polar organic materials from the current
transients. Calculation of $\sigma_\textrm{dip}$ for several
transport materials with varying concentration of polar dopants
gives concentration dependences that are in reasonable agreement
with theoretical predictions. A possible solution of the puzzle
concerning the disorder effect on the mobility temperature
dependence has been suggested.


\section{Introduction}
Modern paradigm of the hopping charge transport in disordered
organic materials (molecularly doped polymers, low-molecular
weight organic glasses, amorphous polymers) states that the most
important factor governing the behavior of a charge carrier is
energetic disorder \cite{Bassler:15,Pope:1328}. It is generally
accepted that hopping of a charge carrier in random energy
landscape $U(\vec{r})$ describes very well all major features
observed in transport experiments. Crucial ingredients of the most
successful realization of this idea includes the Gaussian density
of localized transport states
\begin{equation}
g(U)=\frac{N_0}{(2\pi\sigma^2)^{1/2}}\exp\left(-\frac{U^2}{2\sigma^2}\right),
\label{Gauss}
\end{equation}
and Miller-Abraham hopping rate \cite{Miller:745}; here $N_0$ is a
total concentration of transport sites and $\sigma$ is rms
disorder. If we assume no spatial correlation between energies of
different transport sites (a famous Gaussian Disorder Model (GDM)
\cite{Bassler:15}), then the field and temperature dependence of
the carrier drift mobility $\mu$ could be described as
\cite{Bassler:15}
\begin{equation}
\mu =\mu_0\exp\left(-2\alpha
a\right)\exp\left[-\frac{4}{9}\hat{\sigma}^2+C\left(\hat{\sigma}^2-\Sigma^2\right)\sqrt{E}\right],
\hskip10pt \hat{\sigma}=\sigma\beta, \hskip10pt \beta=1/kT,
\label{GDM}
\end{equation}
where $a$ is a lattice scale for the equivalent homogeneous
lattice model with the same concentration of transport sites ($a$
scales as $c^{-1/3}$ with concentration $c$ of transport sites),
$\alpha$ is an inverse localization radius of the wave function of
the transport level, parameter $C$ scales as $a$ and parameter
$\Sigma$ describes the positional disorder (more precisely, if
$\Sigma < 1.5$, then the term $\Sigma^2$ in eq \ref{GDM} should be
replaced by 2.25).

In fact, the Poole-Frenkel (PF) mobility field dependence
$\ln\mu\propto\sqrt{E}$,usually observed in experiments, could be
described by the GDM only in rather narrow field range around
$E\approx 1\times 10^6$ V/cm. This limitation was naturally
removed in the more advanced model of correlated disorder
\cite{Novikov:14573,Gartstein:351,Dunlap:542,Novikov:4472}.
Indeed, one of the basic assumptions of the GDM, i.e. the
assumption of the absence of spatial correlations in the random
energy landscape proves to be spectacularly wrong in organic
materials. In these materials interaction of a charge carrier with
randomly located and oriented dipoles (the model of dipolar glass
\cite{Novikov:14573}) or quadrupoles (the model of quadrupolar glass
\cite{Novikov:89}) generates highly correlated energy landscape
with correlation function $C(r)=\left<U(\vec{r})U(0)\right>$
decaying as $1/r$ for the dipolar glass
\cite{Novikov:14573,Dunlap:542} and as $1/r^3$ for the quadrupolar
glass \cite{Novikov:89}. The model of dipolar glass (DG) is a good
candidate to describe transport properties of polar disordered
organic materials, while the model of quadrupolar glass (QG)
naturally describes non-polar materials having zero dipolar moment
but nonzero quadrupolar moment; quadrupolar moments are
sufficiently high for transport molecules having local polar
groups with compensating dipolar moments \cite{Novikov:89}.

Correlated landscape in the dipolar glass naturally produces the
Poole-Frenkel mobility field dependence, as it was shown in one
dimensional (1D) model of charge transport, where correlation
function with power-law decay $C(r)\propto 1/r^n$ leads to the
mobility field dependence $\ln\mu\propto E^{n/(n+1)}$
\cite{Dunlap:542}. This conclusion has been confirmed by the 3D
Monte Carlo simulation \cite{Novikov:4472}, where the mobility
obeys the relation
\begin{equation}
\mu =\mu_0\exp\left(-2\alpha
a\right)\exp\left[-\frac{9}{25}\hat{\sigma}^2+C\left(\hat{\sigma}^{3/2}-\Gamma\right)\sqrt{eaE/\sigma}\right],
\label{DGmu}
\end{equation}
with $C\approx 0.78$ and $\Gamma\approx 2$ for the case of totally
filled lattice. In striking contrast with the GDM, the model
of dipolar glass (sometimes called the correlated disorder model)
demonstrates the Poole-Frenkel field dependence in the field range from
$10^4$ V/cm to $10^6$ V/cm, in good agreement with experimental
data. For
 nonpolar materials the QG model gives $\ln\mu\propto E^{3/4}$
\cite{Novikov:89}, but this dependence is hardly distinguishable
from the PF dependence in rather narrow field range typical for
time-of-flight (TOF) data in nonpolar materials
\cite{Borsenberger:9,Borsenberger:555,Borsenberger:233}.

Recent paper by Schein and Tyutnev \cite{Schein:7295} put forward a
challenge to this picture. They analyzed a vast set of data on the
temperature dependence of the mobility, extrapolated to the zero
electric field (in short, mobility temperature dependence).
According to eqs \ref{GDM} and \ref{DGmu}, $\ln\mu(E\rightarrow
0)\propto -\hat{\sigma}^2$, giving us an obvious opportunity to
calculate $\sigma$ from the TOF data. If we assume that different
sources of energetic disorder are independent, then the total
$\sigma^2$ is a sum of contributions from the individual sources
\begin{equation}
\sigma^2=\sum_i \sigma^2_i. \label{sum_s2}
\end{equation}
One of the most important sources of energetic disorder in polar
organic materials is a dipolar disorder, properly described by the
DG model. For a lattice version of the DG model on a simple cubic
lattice
\begin{equation}
\sigma^2_{\textrm{dip}}=5.53 \frac{e^2p^2 c}{\varepsilon^2 a^4},
\label{s2_dip}
\end{equation}
here $c$ is a fraction of sites, occupied by dipoles, $p$ is a
dipole moment, and $\varepsilon$ is a dielectric constant of the
medium \cite{Novikov:877e,Young:435}.

We should expect that other sources provide contributions to the
sum in eq \ref{sum_s2}, e.g. charge-quadrupolar interactions or
van-der-Waals (charge-induced dipole) interactions. Nonetheless,
if we consider highly polar transport dopants (with $p= 3-5$ D) in
nonpolar polymer binder, e.g. polystyrene (PS), then we could
reasonably expect that the dipolar term (eq \ref{s2_dip}) gives a
dominant  contribution to the total sum, so $\sigma$ should
decrease with the decrease of the concentration of transport
dopant $c$. Yet in   \cite{Schein:7295} it was noted that
experimental data demonstrate approximately \textit{constant}
magnitude of $\sigma$ for $c$ varying from 10\% to 70\%. The only
way to provide an agreement between eq \ref{s2_dip} and
experimental observations seems to suggest a contribution to eq
\ref{sum_s2} which increases with the decrease of $c$. Although
such contributions do really exist (for example,
charge-quadrupolar or van-der-Waals interactions of charge carrier
with polystyrene units), their contributions are expected to be much
smaller than the dominant dipolar contribution. According to the
analysis of Schein and Tyutnev, the situation is very general. For
this reason they concluded that the most possible reason for the
mobility temperature dependence is a contribution from some
\textit{intramolecular} mechanism. A potential reader of their
paper may ask an unavoidable question: where is the disorder? Why
does not it manifest itself in the temperature dependence of the
mobility? In this paper we are going to discuss this fundamental
problem affecting any possible disorder model of charge carrier
transport in organic materials.

\section{How to calculate $\sigma_{\textrm{dip}}$}

In order to resolve a problem of the mysterious absence of the
disorder effect, let us start with the examination of the crucial
difference between the GDM and correlated models. We already noted
that the PF behavior cannot be described in any convincing way by
the GDM but arises naturally in the DG model. The most important
transport property of the correlated disorder is a relation
between the spatial decay of the correlation function and the
mobility field dependence \cite{Dunlap:542}. This fundamental
relation offers a possibility to calculate a contribution from the
dipolar disorder to the total $\sigma$ \textit{directly from the
TOF data}.

This can be done because according to the correlated disorder
formalism (the simplest tool here is a 1D transport
model developed in   \cite{Dunlap:542}) the most important
contribution to the mobility field dependence for moderate field
is provided by the most correlated component of the disorder, i.e.
the component with the slowest decay of the correlation function
$C(r)$. For example, for the algebraic correlation
function $C(r)=A\sigma^2\left(a/r\right)^n$ in the most important
case of strong disorder $\sigma\beta\gg 1$ the 1D model gives
\begin{equation}
\ln\mu/\mu_0=G(T,E)\approx-2\alpha
a-\sigma^2\beta^2+\left(1+\frac{1}{n}\right)
\sigma\beta\left(An\sigma\beta\right)^{\frac{1}{n+1}}\left(\frac{eaE}{\sigma}\right)^{\frac{n}{n+1}}.
\label{1D}
\end{equation}
As we already noted, $n=1$ for the dipolar disorder
 and $n=3$ for the quadrupolar disorder; in both cases a dimensionless constant $A\simeq 1$
\cite{Novikov:14573,Novikov:89}. Formally, the GDM case may be
considered as the limit $n\rightarrow \infty$. Mobility field
dependence can be estimated by the derivative
\begin{equation}
\frac{\partial G}{\partial
E}=\frac{\sigma}{ea}\left(An\sigma\beta\right)^{\frac{1}{n+1}}\left(\frac{\sigma}{eaE}\right)^{\frac{1}{n+1}}.
\label{dSdE}
\end{equation}
Taking into account that $\sigma\beta\gg 1$ and $eaE/\sigma \leq
1$, we see that the right hand side in eq \ref{dSdE} decreases
with the increase of $n$, i.e. with the decrease of the strength of
spatial correlation (contribution of the multiplier
$n^{\frac{1}{n+1}}$ is unimportant). Thus, eq \ref{dSdE} indicates
that the most correlated disorder provides the most important
contribution to the mobility field dependence.

Dominant role of the dipolar disorder could be easily demonstrated
for the case of the two component disorder
\begin{equation}
U(\vec{r})=U_{\textrm{dip}}(\vec{r})+U_n(\vec{r}),
\label{two_terms1}
\end{equation}
where some less correlated independent random energy
$U_n(\vec{r})$ gives an additional contribution to the total
disorder. If the correlation function for this contribution has a
form $C_n(\vec{r})=A_n\sigma^2_n\frac{a^n}{r^n}$ with $n > 1$,
then the 1D model shows that at the low field boundary of the PF
region $ea\beta E\gtrsim
\left(\sigma^2_{\textrm{dip}}\beta^2\right)^{-1}$ the additional
contribution to the mobility field dependence is small, if
\begin{equation}
n\hskip2pt\frac{\sigma^2_n}{\sigma^2_{\textrm{dip}}}\hskip2pt
\frac{1}{\left(\beta^2\sigma^2_{\textrm{dip}}\right)^{n-1}} \ll 1
\label{result_final1}
\end{equation}
(see Appendix). In many organic materials $\beta\sigma_{\textrm{dip}}\gg 1$ and dipolar contribution dominates the mobility field
dependence even if $\sigma_n\simeq\sigma_{\textrm{dip}}$.

\begin{figure}[tbh]
\begin{minipage}[c]{0.5\linewidth}
\begin{center}
\includegraphics[width=2.9in]{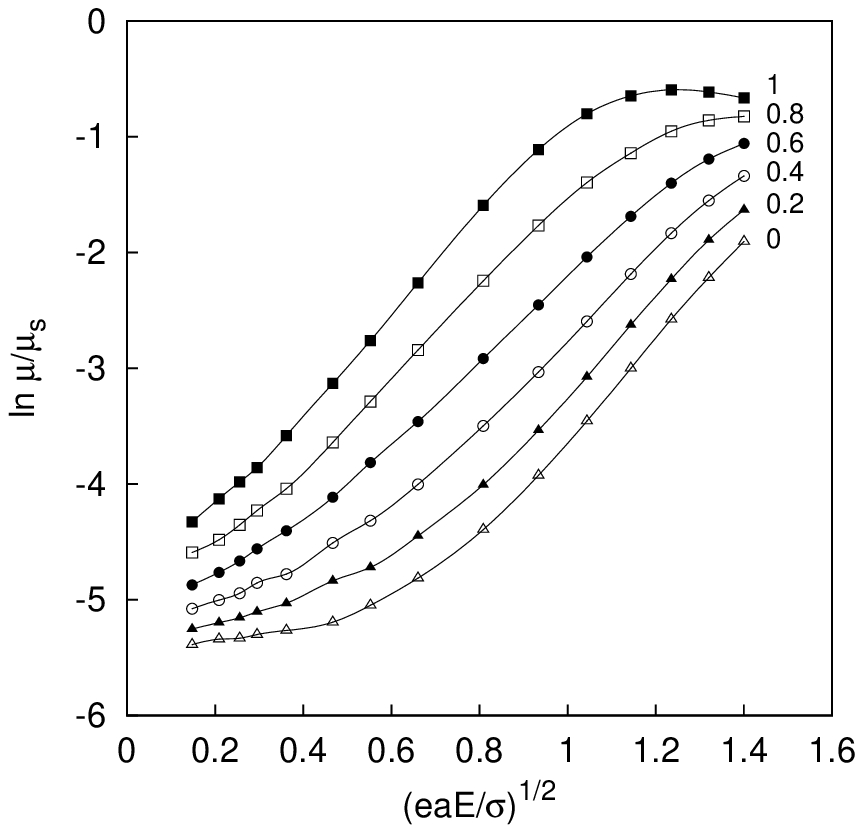}
\end{center}
\end{minipage}
\begin{minipage}[c]{0.5\linewidth}
\begin{center}
\includegraphics[width=2.9in]{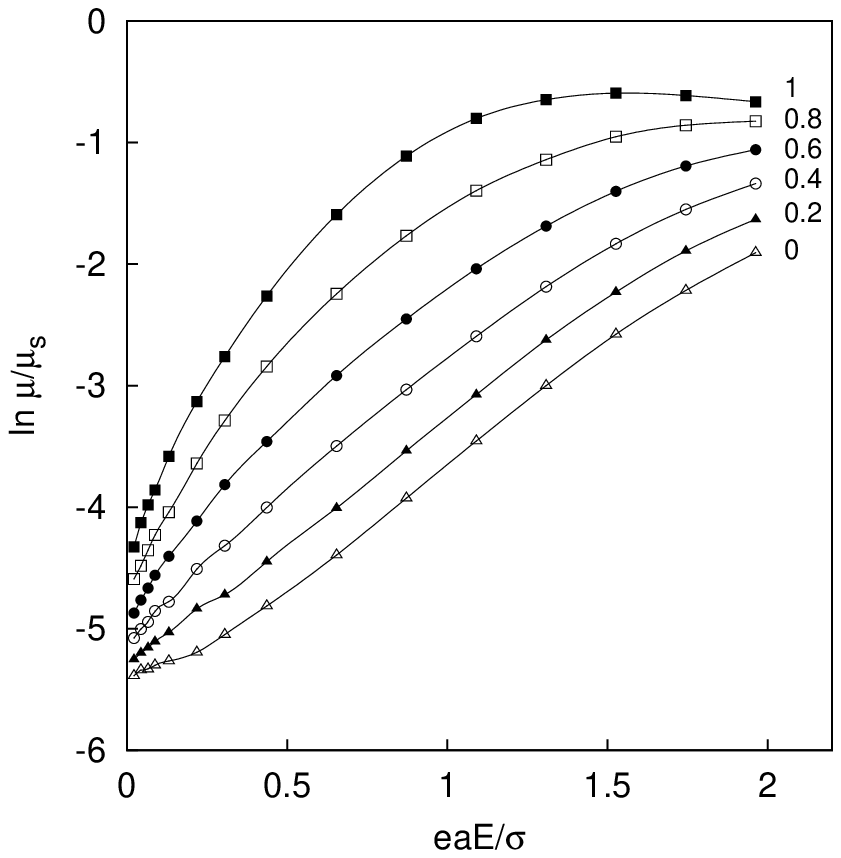}
\end{center}
\end{minipage}
\vskip10pt \hskip120pt a\hskip220pt b \caption{Mobility field
dependences for a mixture of dipolar and non-correlated GDM
disorder for the constant value of the total $\sigma$; here
$\mu_s=\mu_0\exp\left(-2\alpha a\right)$,
$\sigma^2=\sigma^2_{\textrm{dip}}+\sigma^2_{\textrm{GDM}}$, and
the ratio $r=\sigma^2_{\textrm{dip}}/\sigma^2$ is indicated at the
corresponding curve. Transients have been simulated for
$kT/\sigma=0.26$. If $a\approx 1$ nm and $\sigma\approx 0.1$ eV,
then $eaE/\sigma\approx 1$ for $E=1\times 10^6$ V/cm. Note the
transformation of the mobility field dependence for moderate field
from the PF type for $r=1$ to the linear dependence $\ln\mu\propto
E$ for $r=0$. This very dependence is an intrinsic property of the
GDM \cite{Novikov:4472}.} \label{fig_mixed}
\end{figure}

 The
result of the 1D model could be supported by the direct 3D Monte
Carlo simulation of charge carrier transport in the energy
landscape comprised by the mixture of the dipolar and
non-correlated GDM disorder (see Figure \ref{fig_mixed}). In
Figure \ref{fig_mixed}a the decrease of the slope
$S=\frac{\partial G}{\partial E^{1/2}}$ of the mobility field
dependence (or decrease of $\frac{\partial G}{\partial E}$ in
Figure \ref{fig_mixed}b) for $eaE/\sigma \leq 1$ with the decrease
of dipolar contribution $\sigma_{\textrm{dip}}$ to the total
$\sigma$ is clearly visible.

Hence, the temperature dependence of the slope $S(T)$
\begin{equation}
S(T)=\frac{\partial G}{\partial
E^{1/2}}=C\left(\hat{\sigma}^{3/2}-\Gamma\right)\sqrt{ea/\sigma}
\label{S(T)}
\end{equation}
 in polar materials could be used \textit{for a direct
evaluation of} $\sigma_{\textrm{dip}}$. Other sources of disorder,
as well as the contribution from traps \cite{Novikov:2584},
provide much less significant contributions to $S(T)$ for $eaE <
\sigma_{\textrm{dip}}$.  Situation is especially favorable for
highly polar dopants in PS matrix because the dipolar moment of
styrene molecule is only 0.4 D \cite{Borsenberger:11314}, and we
can expect that the quadrupolar moment of styrene is rather low
too (styrene molecule is built only by carbon and hydrogen atoms
and does not contain polar groups). Traps do not affect $S(T)$,
either \cite{Novikov:181}. In future we will use notations
$\sigma_S$ for the rms disorder, calculated from $S(T)$, and
$\sigma_T$ for the corresponding rms disorder, calculated from the
temperature dependence of $\mu(E\rightarrow 0)$.

\begin{figure}[tbh]
\begin{center}
\includegraphics[width=3.5in]{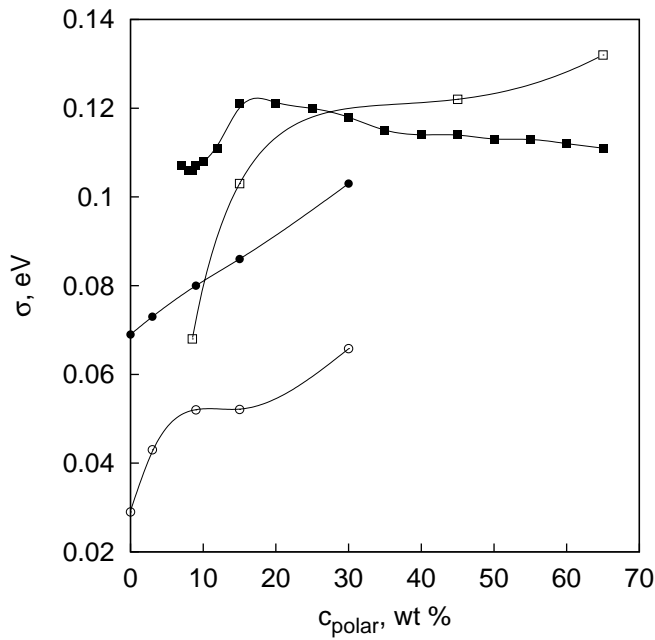}
\end{center}
\caption{Dependence of $\sigma$ on the concentration of polar
constituents \cite{Borsenberger:166,Borsenberger:11314}:
$\sigma_S$ (empty symbols) and $\sigma_T$ (filled symbols),
correspondingly. Squares correspond to the system DEH:PS
\cite{Borsenberger:166}, circles correspond to the system with
copolymer PS-BA as a binder \cite{Borsenberger:11314}, and lines
are shown as a guide for an eye. Data from the paper  \cite{Borsenberger:185} are not shown
to avoid overcrowding of the figure.} \label{fig_c_dip}
\end{figure}

\begin{figure}[tbh]
\begin{center}
\includegraphics[width=3.5in]{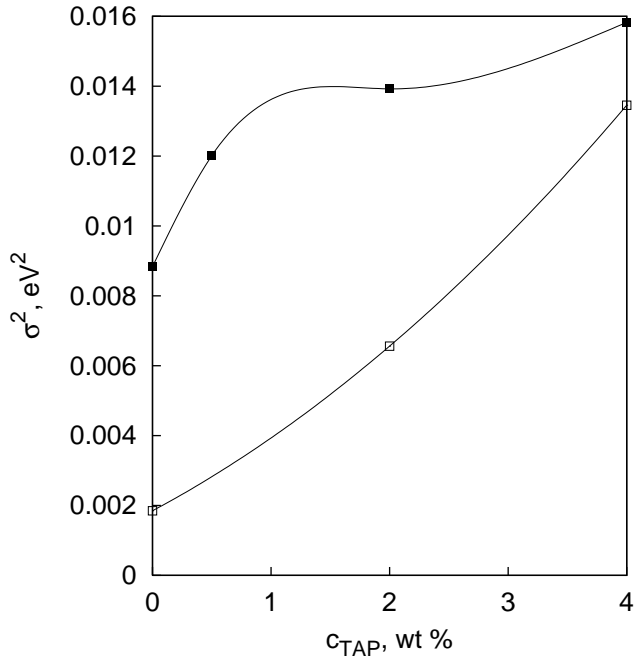}
\end{center}
\caption{Dependence of $\sigma^2$ on the concentration of inert
polar dopant TAP \cite{Young:6290}: $\sigma^2_S$ (empty squares)
and $\sigma^2_T$ (filled squares), correspondingly. Lines are
shown as a guide for an eye.} \label{fig_c_TAP}
\end{figure}

Information we needed is pretty rare in literature and we
found only four papers where the data for $S(T)$ have been provided for different
concentrations of polar constituents
\cite{Borsenberger:11314,Borsenberger:166,Young:6290,Borsenberger:185}.
Fortunately, these papers describe very different situations. Two papers consider dependence of the mobility on
the concentration of highly polar transport dopants DEH
(4-diethylaminobenzaldehyde diphenylhydrazon, $p=3.61$ D)
\cite{Borsenberger:166} and TPM-E (triphenylmethane derivative,
$p=2.1$ D) \cite{Borsenberger:185} in PS matrix. The third paper describes dependence of the mobility on
the concentration of polar units in the copolymer of styrene with
polar monomer BA (butylacrilate, $p=1.52$ D)
\cite{Borsenberger:11314}, serving as a polymer binder, and the
last paper considers an effect of the inert polar dopant TAP
(\textit{t}-amylphthalonitrile, $p=6.6$ D) \cite{Young:6290} on
the charge transport in weakly polar transport material. In
the two last cases concentration of the transport dopant was fixed. In all
cases calculated $\sigma_{\textrm{dip}}$ demonstrates the same
expected trend (see Figures \ref{fig_c_dip} and \ref{fig_c_TAP}):
it decreases with the decrease of the concentration of polar
constituents with the expected magnitude of the decrease (apart from the case of TPM-E:PS, where the decrease is rather small).

It is worth noting the almost linear dependence of $\sigma_S^2$
on $c$ for the transport material with small concentration of
polar dopant TAP, in good agreement with eq \ref{s2_dip} (Figure
\ref{fig_c_TAP}). We should expect that the model of ideal dipolar
glass without any orientational correlation between dipoles is
best suited for a description of this very situation: low
concentration of polar dopant in amorphous material. Concentration
dependencies of $\sigma_S$ in Figure \ref{fig_c_dip} do not obey
eq \ref{s2_dip}. This deviation may reflect a partial
orientational ordering of dipolar molecules for high concentration
of dipoles or a defect of the particular method of the mobility
calculation, discussed in detail in the next section.

 We would like to
emphasize that the very scarcity of the published data on the
temperature dependence of $S$ could be considered as a strong
argument in favor of general agreement between predictions of the
correlated disorder model and experiment. Four papers
\cite{Borsenberger:11314,Borsenberger:166,Young:6290,Borsenberger:185}
may be regarded as four random picks chosen from the vast variety
of the TOF data. Remarkably, these four random picks immediately
demonstrate reasonable agreement with theoretical predictions.

\begin{figure}[tbh]
\begin{center}
\includegraphics[width=3.5in]{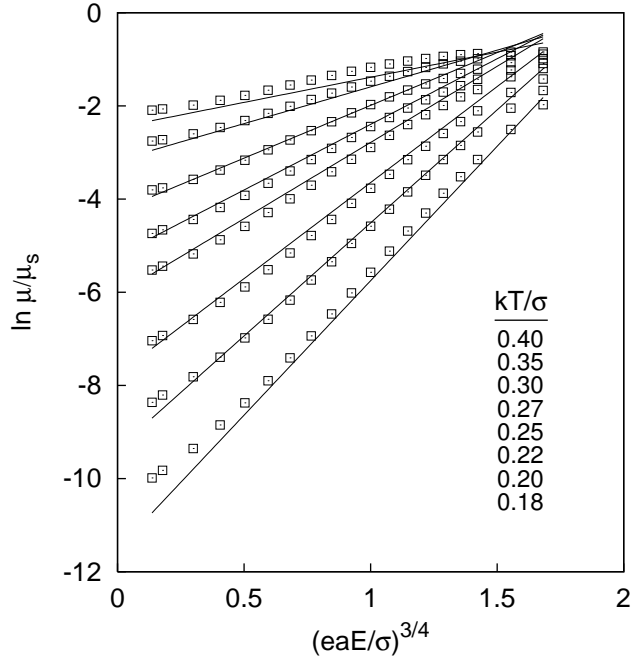}
\end{center}
\caption{Mobility field dependence in the QG model for different
values of $kT/\sigma$ (from the top curve downward); straight lines
indicate the fit for eq \ref{QGmu}. Plot of the simulation data in the usual PF presentation $\ln\mu$ vs $E^{1/2}$ demonstrates much stronger deviation from the linearity in the weak field region. } \label{fig_QG}
\end{figure}

If the dipole moments of all constituents of the transport layer
are low, then the DG model is clearly inadequate for the
description of the charge transport. The most promising candidate
in that case is the QG model. Here we provide only a very brief
summary of the Monte Carlo simulation results for the QG model,
more complete report will be published elsewhere. General approach
is almost identical to the one described in \cite{Novikov:4472}.
Results of the simulation (see Figure \ref{fig_QG}) could be
summarized as
\begin{equation}
\mu =\mu_0\exp\left(-2\alpha
a\right)\exp\left[-0.37\hat{\sigma}^2+C_Q\left(\hat{\sigma}^{5/4}-\Gamma_Q\right)\left(eaE/\sigma\right)^{3/4}\right],
\label{QGmu}
\end{equation}
with $C_Q=0.87$ and $\Gamma_Q=1.91$. In nonpolar organic materials
mobility field curves should be fitted to eq \ref{QGmu}, and the
corresponding quadrupolar $\sigma$ should be estimated from the
temperature dependence of the slope of the dependence $\ln\mu$ vs
$E^{3/4}$.

\section{Dangers of low field mobility}

In the previous section it was demonstrated that dipolar energetic
disorder in organic materials can be directly evaluated from the
TOF data. Result of this evaluation demonstrates a reasonable
behavior of $\sigma_{\textrm{dip}}$ with respect to concentration
of polar constituents. Magnitude of $\sigma_{\textrm{dip}}$ (and
its variation with concentration) again is in reasonable agreement
with the theoretical estimation by eq \ref{s2_dip}. Now we can
reformulate the crucial question raised in the Introduction in a
different way: why does the energetic disorder manifest itself in
the mobility field dependence, but not in the mobility temperature
dependence? Why the difference?

Discussion in the previous section unambiguously demonstrates that
 disorder does affect hopping transport of charge carriers in
 organic materials. We cannot consider as a possible alternative
explanation an assumption that some totally unknown mechanism is
responsible for the PF field dependence and, at the same time, it
just by accident provides the same dependence of transport
parameters on the concentration of polar dopants as a well known
mechanism of the correlated dipolar disorder.

If disorder does affect charge transport, then there is no
possibility to avoid a corresponding term proportional to
$\sigma^2_{\textrm{dip}}$ in the mobility temperature dependence.
According to the estimations of eqs \ref{s2_dip} and \ref{DGmu},
this term is not negligible but quite comparable to the total
$\sigma^2$. Hence, only two possibilities are feasible: either
some compensating mechanism provides a contribution to the total
$\sigma$ (more rigorously, to the slope of the mobility
temperature dependence) or some defects of the experimental
procedure masquerade the true dependence of $\sigma$ on $c$. Let
us consider both possibilities.

\subsection{Possible physical compensating mechanisms}

In the previous sections we demonstrated that $S(T)$ gives a
robust estimation for the magnitude of $\sigma_\textrm{dip}$.
Temperature dependence of the low field mobility is not so easily
tractable. A host of various contributions adds to the resultant
dependence: charge-dipole, charge-quadrupole, van der Waals
(charge-induced dipole) interactions, polaron contribution of a
simple Arrhenius type \cite{Parris:126601}, and the similar
contribution from traps. Reliable separation of the individual
contributions to the resultant temperature dependence
 is impossible. As it was already mentioned, some contributions
could provide compensating effects, for example contributions
originated from the carrier interactions with polymer binder,
though the magnitude of the effect is expected to be insufficient
for a total compensation in the case of highly polar transport
dopant in the PS binder. There is another possibility for
compensation because for a low concentration of the dopant charge
transport should become more sensitive to the energetic disorder:
exponential decrease of the hopping rate for large average
distances between dopant molecules results in carrier hopping to
the nearest neighbor without regards to its energy. This means
that the effective $\sigma_T$ should increase with the decrease of
$c$, but this effect is hardly very important for $c > 10-15\%$.

Effect of impurities (possible traps) deserves a special discussion. First of all, traps
do affect effective magnitude of $\sigma_T$, at least the one
estimated using the GDM \cite{Wolf:259}. Effect of impurities could be responsible for different values of $\sigma_T$ observed for the same material. For example, dependence
of $\sigma_T$ on $c$ for the same system DEH:PS was found to be
different in different papers \cite{Borsenberger:166,Schein:773}.
In the earlier paper \cite{Schein:773} a constant value
$\sigma_T\approx 0.13$ eV was found. In the later paper
\cite{Borsenberger:166} more rigorous purifying procedures were used and very careful preparation procedure was undertaken. As a result, variation $\simeq 15\%$ in the magnitude
of $\sigma_T$ over studied concentration range has been found (see
the corresponding curve in Figure \ref{fig_c_dip}). This is
certainly beyond the experimental error range. For the same
concentration of DEH, magnitude of $\sigma_T$, reported in Refs.
\cite{Borsenberger:166} and \cite{Schein:773}, differs up to
20-25\%. This means that the slope of the mobility temperature
dependence could differ up to 40\% for different procedures of
material purification and sample preparation. In
\cite{Schein:7295} the result of   \cite{Borsenberger:166} is
mistakenly regarded as corroborating the conception of
intramolecular origin of $\sigma_T$, while in fact comparison of
the results of Refs. \cite{Borsenberger:166} and \cite{Schein:773}
suggests that more careful
examination of the trap effect on $\sigma_T$, calculated from the
TOF data, is needed.

Some details of the preparation procedure, exploited in
\cite{Borsenberger:166}, are unusual. For example, coating of the transport layer was performed under the red
light to avoid photochemical creation of traps. Quite probably,
this very careful preparation procedure gave a possibility to
avoid excessive amount of traps and obtain more reliable values of
$\sigma_T$.

Note, that polymers are difficult to purify; to some extent, all of them contain various impurities,
possibly serving as traps. Analysis of charge transport in
trap-containing materials shows that increase in trap
concentration could be interpreted as an increase of the effective
$\sigma_T$ \cite{Borsenberger:1945}. In molecularly doped polymers
decrease of the dopant concentration means a simultaneous increase
of the polymer concentration and, thus, possible increase in
concentration of traps. This could provide an additional
compensating mechanism.

\subsection{Deficiency of the mobility calculation}

So far we considered a possible compensation that is relayted to
some intrinsic physical mechanism. Yet, quite probably, the most
important reason for the compensation is a peculiar procedure for
the calculation of the drift mobility, employed in the majority of
experimental papers.

There are two most frequently used procedures for the calculation
of $\mu=L/Et$ from the TOF data in double linear current vs time
presentation: 1) calculation that uses time $t_i$ determined by
the intersection of asymptotes to the plateau and trailing edge of
the transient and 2) calculation that uses time $t_{1/2}$ for
current to reach half of its plateau value; here $L$ is a
thickness of the transport layer. The first method is a method of
choice for most papers. One can find in literature the
statement that the difference between two procedures is not very
important for determination of the temperature and field
dependence of the mobility \cite{Borsenberger:79,Van:31}. This is
not true.

\begin{figure}[tbh]
\begin{center}
\includegraphics[width=3.5in]{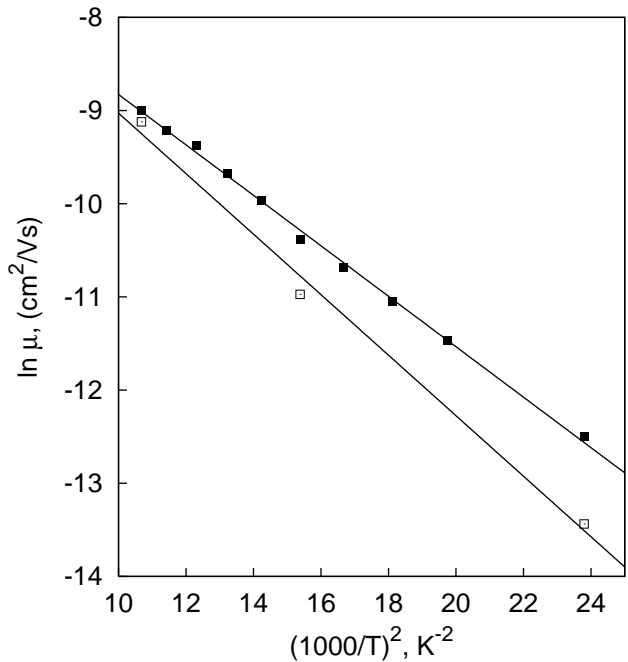}
\end{center}
\caption{Mobility field dependence for two methods of the mobility
calculation: $\mu_i$ (filled squares) and $\mu_{1/2}$ (empty
squares) \cite{Borsenberger:435}. Straight lines indicate best
fits.} \label{fig_T}
\end{figure}

Calculation, based on $t_i$, significantly overestimates
contribution of fast carriers. Deficiency of this calculation is
especially evident in the case of materials with traps. In
trap-free materials the use of $\mu_i$ leads only to the change of
the slope of the mobility field dependence. In the presence of
traps the use of $\mu_i$ sometimes distorts even the functional
type of the mobility field dependence, bringing it to the linear
one $\ln\mu_i\propto E$. This behavior was observed in
experiment \cite{Veres:377}, but it is an artifact, directly
related to the inferior method of the mobility calculation
\cite{Novikov:444}. Procedure that uses $t_{1/2}$ is much more
reliable and generally agrees well with the standard definition of
the mobility as $\mu=v/E$, where $v$ is an average carrier
velocity.

Unfortunately, errors, related to the use of $t_i$, are not
limited to the mobility field dependence, they also affect the
mobility temperature dependence. As an illustration we provide
Figure \ref{fig_T}, where the temperature dependence of $\mu_i$
and $\mu_{1/2}$ for tri-{\it p}-tolylamine doped polystyrene
\cite{Borsenberger:435} is shown. Slopes of the dependencies
differ by approximately 20\%, thus the values of $\sigma_T$,
calculated from the slopes, do differ too. Similar difference was found for the data presented in \cite{Bassler:1677}. Again,
as in the case of the temperature dependence of $S$, the data for
raw transients (note that here we need transients for several
values of $T$) is very difficult to find in literature.

It is very important to understand the full significance of the
described fact. If we calculate $\sigma_T$ from the slope of the
mobility temperature dependence for different values of $c$ and
use the mobility $\mu_i$, then \textit{there is absolutely no
guarantee} that its relative difference from the value of
$\sigma_T$, calculated from $\mu_{1/2}$, remains constant in the
whole range of concentration.  Remember, $\mu_{1/2}$ is a much
better approximation for a true mobility. Hence, the obtained concentration
dependence of $\sigma_T$  is certainly not very reliable and, quite
probably, has considerable errors.

\subsection{Non-dispersive or dispersive transport?}

There is another aspect of the mobility calculation procedure that
can be a possible source of errors. It was claimed in
\cite{Schein:7295} that for all cases used in the analysis of
charge transport, transport regimes are essentially
non-dispersive. This is not true. For example, Borsenberger
\textit{et al.} explicitly emphasized that transients for highly
polar dopant DTNA (di-{\it p}-tolyl-{\it p}-nitrophenilamine,
$p=5.78$ D) in PS are highly dispersive and no plateau was
observed in the whole field, temperature and concentration range;
sometimes the transit time has been even determined by the
intersection of the asymptotes in double logarithmic current vs
time representation \cite{Borsenberger:171}. For this very reason
$\sigma_T$ was even calculated by the formula
\begin{equation}\label{dispersive}
    \ln\mu/\mu_0\approx -2\alpha a -\frac{1}{4}\widehat{\sigma}^2
\end{equation}
that differs from eq \ref{GDM}. It was suggested by B{\"a}ssler
and Borsenberger \cite{Bassler:763} for the treatment
of low field mobility in the case
 of dispersive transport.
Particular material DTNA:PS could be an exceptional case in the
sense that in the whole range of all relevant parameters $c$, $T$,
and $E$ transients demonstrate no visible plateau, but for low
temperature ($T \le 230$ K) this kind of transients is generally a
rule and not an exception
\cite{Bassler:763,Borsenberger:12145,Borsenberger:4289}; how the
transit time $t_i$ (or $t_{1/2}$) could be reliably calculated in
double linear plot is not very clear in this situation.

In fact, sometimes the problem of the choice between
non-dispersive and dispersive transport regime is even more tricky
than described in the previous paragraph. For example, let us
consider again the case of DTNA-doped PS. In the original paper
\cite{Borsenberger:171} transients were considered as
dispersive because no visible plateau was observed. Yet it is
very well known that non-dispersive transport (described by the
usual diffusion equation with well defined $v$ and diffusion
coefficient $D$) can routinely produce the same kind of transients
if $v L/D \le 5-10$. Such transients have no well defined plateau
but could be easily discriminated from the true dispersive
transients by their asymptotes for $t\rightarrow \infty$. True
dispersive transport transients obey the law $I(t)\propto
t^{-1-\alpha}$ \cite{Scher:2455}, while for a normal diffusion
 $I(t)\propto \exp\left(-v^2t/4D\right)$, putting aside an unimportant power-law factor.

\begin{figure}[tbh]
\begin{center}
\includegraphics[width=3.5in]{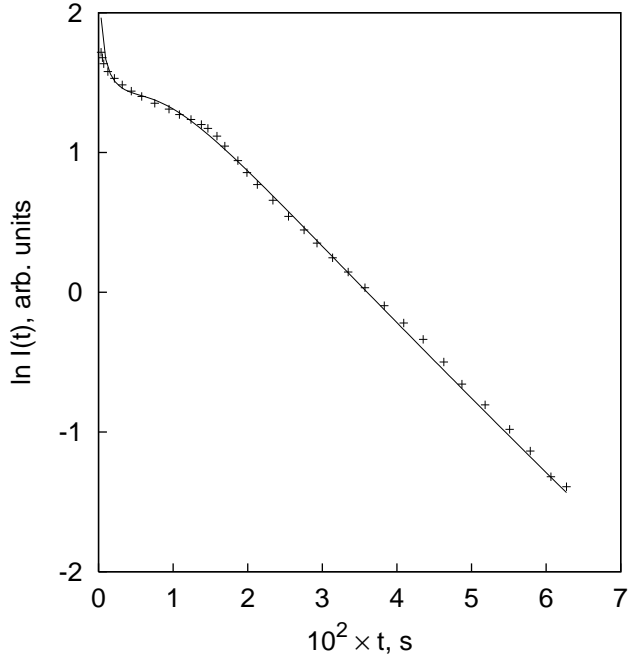}
\end{center}
\caption{Transport transient from  Figure 2 of
\cite{Borsenberger:171} (crosses) and the best fit for eq
\ref{hirao} (line). Note that $\ln I(t)\propto t$ for $t \gg
L/v$.} \label{DTNA}
\end{figure}

This test immediately demonstrates that at least some transients,
provided in   \cite{Borsenberger:171}, \textit{are not dispersive}: the transient can be fitted with reasonable accuracy by the
solution of the diffusion equation
\cite{Hirao:1787,Hirao:4755,Nishizawa:L250}
\begin{equation}
I(t)\propto\sqrt{\frac{D}{\pi t}} \left[\exp \left(-\frac{v^2
t}{4D}\right)- \exp\left(-\frac{(L-vt)^2}{4Dt}\right)\right]
+v\left[{\rm erf}\left(\frac{L-vt}{\sqrt{4Dt}}\right)+ {\rm
erf}\left(\frac{vt}{\sqrt{4Dt}}\right)\right], \label{hirao}
\end{equation}
where $\textrm{erf}(x)$ is the error function, and the large time
asymptotic is of the exponential kind (see Figure \ref{DTNA}). We
can safely conclude that in some cases charge transport in DTNA:PS
occurs in the non-dispersive regime, and yet it was always analyzed using
eq \ref{dispersive}, suitable only for highly dispersive regime!
All this confusion means that the true magnitude of $\sigma_T$ and
its dependence on $c$ are very different from provided in
\cite{Borsenberger:171}.
 One can learn a very important lesson from this
example: in order to calculate mobility from transients without
plateau the very first step should be an attempt to fit the
transient to eq \ref{hirao}.

One can conclude that the data, discussed in  \cite{Schein:7295},
had been obtained both for non-dispersive and dispersive regimes.
Even worse, in some cases a non-dispersive mobility had been
treated as a dispersive one, leading to the patently wrong
calculation of $\sigma_T$. We believe that in this situation it is impossible to reveal the true dependence of the rms disorder on the concentration of polar constituents.

\subsection{Possible hints for improper treatment of TOF data}

Discussion of possible difficulties relevant to the analysis of
TOF transients and evaluation of the disorder parameters from the
TOF data should be finished with one additional remark. Although
current transients are of primary importance for a detection of
 possible deficiencies of the mobility calculation procedure,
sometimes the hints for the improper treatment of the transport
data could be obtained directly from $\mu(E)$ or $\mu(T)$ curves.

One very well known example is an artifact of the mobility
decrease with the increase of $E$ in weak field region. Earlier
this phenomenon was attributed to the effect of field induced
traps \cite{Bassler:15}, though this explanation for weak field
region always looks pretty suspicious. Later it was found that the
true reason for this strange behavior is a deficiency in the
mobility calculation procedure, i.e. the use
 of $\mu_i$ \cite{Hirao:1787,Hirao:4755,Baranovskii:1644}. If fitting of the transients
to eq \ref{hirao} is used, then the mobility curve becomes
monotonously increasing with $E$. The reason for the unexpected
increase of $\mu$ in weak field region is a trivial contribution
from the normal diffusion, described by the diffusion coefficient
$D$. It is worth noting that the fitting procedure, suggested by
Hirao \textit{et al.} \cite{Hirao:1787,Hirao:4755}, produces an
additional important result: it removes the spurious dependence of
$\mu$ on the thickness of the transport layer $L$ ($D$ was also
found to be independent of $L$).

Artifact of the mobility field dependence in weak field is
directly related to the main subject of our consideration.
Mobility that decays with the increase of $E$ was reported for
transport material TAPC:PC
(1,1-bis-(di-4-tolylaminophenyl)cyclohexane in polycarbonate)
doped with small amounts of inert polar dopants o-, m-, and p-DNB
(isomers of dinitrobenzene) \cite{Borsenberger:291}. For m-DNB and
p-DNB the mobility decays with the increase of $E$ in the whole
field range (up to $E\simeq 1\times 10^6$ V/cm). Such behavior for
room (or lower) temperature is extremely unusual and, very
probably, indicates an improper method of the mobility calculation
from the TOF data. Correspondingly, magnitude of $\sigma_T$,
provided in  \cite{Borsenberger:291} and extensively discussed by
Schein and Tyutnev \cite{Schein:7295}, is hardly very reliable.

Highly unusual value of some relevant parameter, obtained from the
interpretation of TOF data, could serve as another alarm signal.
As an example of such alarm signal we consider an unusual value of
the radius of the localization $R_0$ of the wave function of
transport site, obtained for DEH:PS \cite{Schein:773}. It was
found that $R_0=1/\alpha=4.8$ A, while for all other transport
materials $R_0$ typically falls in the range $1-2.5$ A
\cite{Borsenberger:6263,Borsenberger:233a,Borsenberger:5283,Kitamura:821}.
We believe that this is another indication that something goes
wrong with the mobility calculation; again, the
use of $\mu_i$ is a probable culprit.

\section{Checklist for the reliable mobility calculation and subsequent evaluation of the disorder parameters}

In this section we would like to emphasize principal details of
the  more safe and reliable procedure for the analysis of the TOF
data in order to elucidate disorder parameters.

\begin{itemize}
\item First of all, wherever possible, transport should be
analyzed for a non-dispersive regime only. In this regime
transport parameters are more sensitive for peculiarities of the
random energy landscape (comparison of eqs \ref{GDM} and
\ref{dispersive} supports this statement for the GDM). \item If a
well defined plateau of the transient is observed, then the drift
mobility may be calculated using $t_{1/2}$. Use of $t_i$ should be
completely eliminated from the experimental practice. \item
If there is no well defined plateau, then the transient should be
fitted to eq \ref{hirao} according to the procedure, suggested
by Hirao \textit{et al.} \cite{Hirao:1787}. Only in the case, when
the quality of the fit is clearly inadequate, transport should be
analyzed in the framework of the dispersive regime. \item For the
case of dipolar disorder, $\sigma_{\textrm{dip}}$ should be
calculated from the temperature dependence of the slope of the
mobility field dependence $S(T)=\frac{\partial \ln \mu}{\partial
E^{1/2}}$ for moderate field using eq \ref{DGmu}. The use of the
DG model instead of the GDM is important, because the very essence
of this approach is based on the transport properties of
correlated energy landscape. This method is almost completely
insensitive to the contribution from traps, polaron effects, and
van der Waals interaction. The quadrupolar contribution probably
could not be totally eliminated in this way, but we may reasonably
expect that it is not dominant for polar materials.\item For
nonpolar materials the magnitude of quadrupolar disorder should be
calculated from the temperature dependence of the corresponding
slope $S_Q(T)=\frac{\partial \ln \mu}{\partial E^{3/4}}$ according
to eq \ref{QGmu}. Unfortunately, description of the mobility field
and temperature dependence for real organic materials with
contributions from different sources of disorder having different
correlation properties cannot be described by any simple unified
formula.
\end{itemize}

Discussion of the dependence of $\sigma_T$ on the concentration of
dipolar dopants and other relevant parameters should be performed
 only after calculation of $\mu$ and $\sigma_S$ in the proper
way. We hope that the realization of this program will provide us
with much more reliable knowledge of the crucial parameters of the
energy landscape in disordered organic materials.

\section{Conclusion}

 This paper is a direct response to the recent review
by Schein and Tyutnev \cite{Schein:7295}. They analyzed a vast set
of available transport data for disordered organic materials and
came to the conclusion that in some mysterious way energetic
disorder is not important for the mobility temperature dependence.
The strongest argument in support of this point of view is the
independence of disorder magnitude $\sigma$ on the concentration
of highly polar transport dopants. This is very unexpected because
the dipolar contribution to the disorder should depend on the
concentration of polar dopants. Schein and Tyutnev suggested that
some intramolecular mechanism is responsible for the mobility
temperature dependence.

We have analyzed essentially the same set of data but in a
different manner and came to a very different conclusion. Our
analysis shows that the dipolar contribution could be directly
calculated from the temperature dependence of the slope $S$ of the
mobility field dependence. Results of such calculations
immediately demonstrate a reasonable agreement between experiment
and theory, though available experimental data is extremely
limited.

We have shown that a very popular method of the mobility calculation
by the time of intersection of asymptotes to the plateau region
and trailing edge of the transient is a possible reason why the
calculations of $\sigma$ give misleading results. A simple yet
accurate procedure of the mobility calculation and subsequent
analysis of the mobility field and temperature dependences for
evaluation of the disorder parameters has been discussed.

We would like to emphasize that a very significant contribution to
the realization of the suggested program could be fulfilled by a
simple recalculation of the mobility from already obtained
transients (if available) and subsequent analysis of the
mobility field dependence according to the lines suggested in the
previous section. Also we would like to note that a direct access
to the raw TOF data (current transients) should significantly
improve our ability to check consistency between experimental data
and various theoretical models and would be considered as God's
gift by all theoreticians. Data for TOF transients could be
provided as supplements to the experimental papers, this
possibility is already provided by many leading scientific
journals.

\section{Acknowledgements}
This work was supported by the ISTC grant 3718 and RFBR grants
05-03-90579-NNS-a and 08-03-00125-a. We are grateful to L. Schein
and A. Tyutnev for giving us the opportunity to read their paper
before publication.

\section*{Appendix. Mixed disorder in 1D model}

Suppose that the total random energy $U(\vec{r})$ is a sum of two
independent terms
\begin{equation}
U(\vec{r})=U_{\textrm{dip}}(\vec{r})+U_n(\vec{r}),
\label{two_terms}
\end{equation}
where $U_{\textrm{dip}}(\vec{r})$ is a dipolar contribution with
the correlation function
$C_{\textrm{dip}}(\vec{r})=A_{\textrm{dip}}\sigma^2_{\textrm{dip}}\frac{a}{r}$,
while the second term provides an additional energetic disorder
with the correlation function
$C_n(\vec{r})=A_n\sigma^2_n\frac{a^n}{r^n}$, and $n > 1$. For
independent disorder contributions the resulting correlation
function is the sum of two terms
\begin{equation}
C(\vec{r})=C_{\textrm{dip}}(\vec{r})+C_n(\vec{r}).
\label{two_C_terms}
\end{equation}
In the 1D model the mobility is
\begin{equation}
\mu=\frac{D_0}{E\int_0^\infty dx \exp\left\{-e\beta E
x+\beta^2\left[C(0)-C(x)\right]\right\}}, \label{1D_mu}
\end{equation}
where $D_0$ is a bare (microscopic) diffusion coefficient
\cite{Dunlap:542,Parris:2803}. For the case of strong disorder
$\beta\sigma\gg 1$ a suitable tool for the calculation of integral
\ref{1D_mu} is a saddle point approximation. For example, eq
\ref{1D} is a result of this approximation. In fact, for a pure
dipolar disorder the PF dependence arises exactly in the case when
this approximation is valid. In the saddle point approximation
\begin{equation}
\mu\propto \exp\left[-C(0)\beta^2+R(x_s)\right], \hskip10pt
R(x)=ex\beta E+C(x)\beta^2, \hskip10pt
C(0)=\sigma^2_{\textrm{dip}}+\sigma^2_n,
\label{mu_saddle}
\end{equation}
and $x_s$ is a solution of the equation
\begin{equation}
\frac{dR(x)}{dx}=e\beta
E-A_{\textrm{dip}}\beta^2\sigma^2_{\textrm{dip}}\frac{a}{x^2}-nA_n\beta^2\sigma^2_n\frac{a^n}{x^{n+1}}=0.
\label{1D_saddle}
\end{equation}
Let us find when the mobility field dependence retains the PF
form. This is the case when the third term in eq \ref{1D_saddle}
is a small correction. If we retain the first two terms in eq
\ref{1D_saddle}, then the solution is
\begin{equation}
x_s^0=aZ=a\left(\frac{A_{\textrm{dip}}\beta^2\sigma^2_{\textrm{dip}}}{e\beta
aE}\right)^{1/2}, \label{1D_saddle_xs0}
\end{equation}
and the third term is a correction if
\begin{equation}
A_{\textrm{dip}}\beta^2\sigma^2_{\textrm{dip}}\frac{a}{(x_s^0)^2}
\gg nA_n\beta^2\sigma^2_n\frac{a^n}{(x_s^0)^{n+1}}, \hskip15pt
\frac{nA_n}{A_{\textrm{dip}}}\hskip2pt\frac{\sigma^2_n}{\sigma^2_{\textrm{dip}}}\hskip2pt\frac{1}{Z^{n-1}}
\ll 1. \label{1D_saddle_corr}
\end{equation}
Saddle point approximation is valid if three conditions are valid
\begin{equation}
-\left(\delta x_s\right)^3\left.\frac{d^3
R}{dx^3}\right|_{x=x_s^0} \ll 1, \hskip20pt \delta x_s \ll x_s^0,
\hskip20pt x_s^0 \gg a, \label{1D_saddle_valid}
\end{equation}
where $\delta x_s=\left(\left.\frac{d^2
R}{dx^2}\right|_{x=x_s^0}\right)^{-1/2}$. The first condition
means that the Gaussian approximation is valid in the vicinity of
the minimum of $R(x)$, the second one means that the
 vicinity of the minimum does not reach the boundary of the integration domain,
 and the third one means that we can safely use power-law asymptotic for the correlation
 functions
$C(x)$. All inequalities could be condensed as
\begin{equation}
1\ll Z \ll \beta^2\sigma^2_{\textrm{dip}}, \label{result_ineq}
\end{equation}
and here and in future we drop all coefficients such as
$3/\sqrt{2}$, $A_{\textrm{dip}}$, or $A_n$. This double inequality
is equivalent to
$\left(\beta^2\sigma^2_{\textrm{dip}}\right)^{-1}\ll ea\beta E \ll
\beta^2\sigma^2_{\textrm{dip}}$. The low field boundary of the PF
region can be estimated as $Z_l\lesssim
\beta^2\sigma^2_{\textrm{dip}}$. Substituting $Z$ with $Z_l$ in eq
\ref{1D_saddle_corr} we obtain eq \ref{result_final1} as a final
necessary condition for the small contribution of the additional
non-dipolar disorder in the low PF region $ea\beta E\gtrsim
\left(\sigma^2_{\textrm{dip}}\beta^2\right)^{-1}$.
 The same estimation could be obtained by
comparison of the dominant contribution to the mobility field
dependence with the additional term $\propto A_n\sigma^2_n$.

\end{document}